%Paper: hep-th/9512024
%From: Micha Berkooz <berkooz@physics.rutgers.edu>
%Date: Tue, 5 Dec 1995 17:28:06 -0500

\input harvmac

\def\np#1#2#3{Nucl. Phys. B{#1} (#2) #3}
\def\pl#1#2#3{Phys. Lett. {#1}B (#2) #3}

\def\cmp#1#2#3{Comm. Math. Phys. {#1} (#2) #3}

%%%%%%%%%%%%%%%%%%
%
%  This inputs the macro package epsf.tex

\ifx\epsfbox\UnDeFiNeD\message{(NO epsf.tex, FIGURES WILL BE IGNORED)}
\def\figin#1{\vskip2in}% blank space instead
\else\message{(FIGURES WILL BE INCLUDED)}\def\figin#1{#1}\fi
\def\ifig#1#2#3{\xdef#1{fig.~\the\figno}
\goodbreak\midinsert\figin{\centerline{#3}}%
\smallskip\centerline{\vbox{\baselineskip12pt
\advance\hsize by -1truein\noindent\footnotefont{\bf Fig.~\the\figno:}
#2}}
\bigskip\endinsert\global\advance\figno by1}

\def\ifigure#1#2#3#4{
\midinsert
\vbox to #4truein{\ifx\figflag\figI
\vfil\centerline{\epsfysize=#4truein\epsfbox{#3}}\fi}
\narrower\narrower\noindent{\footnotefont
{\bf #1:}  #2\par}
\endinsert
}

%**************************************************

\def\mqs#1{q^*_{#1}}

\def\bqs#1{{\tilde q}^*_{\tilde {#1}}}

\def\bqd#1{{\tilde q}^\dagger_{\tilde {#1}}}

\def\sd#1{{{\bf D}_{#1}}}
\def\sdd#1{{{\bf {\bar D}}_{\dot{#1}}}}

\def\kqbs#1#2#3{{\partial K_{#1}\over \partial {\tilde
q}^{*{#2}}_{\tilde{#3}}}}
\def\kqs#1#2#3{{\partial K_{#1}\over \partial q^{*}_{{#2}{#3}}}}

\def\wqs#1#2#3{{\partial W^*_{#1}\over\partial q^{*}_{{#2}{#3}}}}

\Title{hep-th/9512024, RU-95-91}
{\vbox{\centerline{A Comment on Non-Chiral Operators in SQCD and its Dual} }}
\smallskip
\centerline{M. Berkooz \footnote{*}{e-mail: berkooz@physics.rutgers.edu}}
\smallskip
\centerline{Department of Physics and Astronomy}
\centerline{Rutgers University}
\centerline{Piscataway, NJ 08855-0849}
\vskip 2cm
\baselineskip 18pt

\noindent

\vfill

We match a few non chiral operators in the electric and magnetic
descriptions of SQCD, suggesting the first evidence of electric-magnetic
duality outside the chiral ring. Algebraically, these non chiral
operators are a module of the chiral ring. Physically, the suggested
correspondence essentially
transforms certain electric gauge invariant composites containing the
electric field strength into magnetic matter composites.

\vfill

\eject

\nref\sem{N. Seiberg, \np{435}{1995}{129}.}%
\nref\ntkna{K. Intriligator, N. Seiberg, RU-95-48,
IASSNS-HEP-95/70.}%
\nref\multic{A. B. Zamolodchikov,
Sov. J. Nucl. Phys. 44(1986)529-533.}%
\nref\cov{J. Wess, Lecture Notes in Physics 77, Springer, Berlin
1978; P. West, Introduction to Supersymmetry and Supergravity, 2nd
ed. World Scientific.}%
\nref\wssbg{P. Wess, J. Bagger, Supersymmetry and Supergravity,
Princeton University Press.}%
\nref\covb{M. A. Luty, W. Taylor IV, MIT-CTP-2440, hep-th/9506098}%
\nref\nks{D. Kutasov, A. Schwimmer, N. Seiberg, EFI-95-68,
WIS/95/27, RU-95-75, hep-th/9510222.}%
\nref\sucnfa{G. Mack,\cmp{55}{1977}{1}; V.K. Dobrev,
V.B. Petkova, \pl{162}{1985}{127}.}%
\nref\sucnfb{P. Argyres, N. Seiberg, Private Communication.}%
\nref\vafaa{W. Lerche, C. Vafa, N.P. Warner, \np{324}{1989}{427}.}%
\nref\kts{D. Kutasov, \np{442}{1995}{447};
D. Kutasov, A. Schwimmer, \pl{354}{1995}{315}.}%
\nref\otr{O. Aharony, J. Sonnenschein, S. Yankielowicz,
\np{449}{1995}{509};
K. Intriligator, R.G. Leigh, M.J. Strassler,
hep-th/9506148; K. Intriligator, \np{448}{1995}{187}.}%
\nref\bnks{T. Banks, A. Zaks, \np{196}{1982}{189}}%

\newsec{Introduction}

There are two sets of operators whose status in N=1 electric-magnetic
duality \sem\ (for a recent review see \ntkna) is unclear.
The first are non chiral operators and the second are chiral operators
of the form, in say SQCD, ${\widetilde Q}^T W_{\alpha} W_{\beta}....Q $. The
first class constitutes most of the operators in the theory and their
mapping under duality is
important if one really wishes to compute a physical process in the dual
theory. These operators, however,
are under much less control then chiral operators and their
transformation under duality is unclear, leaving the nagging,
albeit implausible, worry that this is not an equivalence of the full
theory. The second
class are trivially included in the first class as they are their descendants.
(for example, ${\widetilde Q}^T W_{\alpha} Q$ is obtained from
${\widetilde Q}^T
\sd{\alpha} Q$ (where $\sd{\alpha}$ is the covariant derivative in
superspace) by applying ${\bar D}^2$, and rewriting the constrained
super-connection in terms of $W_{\alpha}$). They are, however, of immediate
relevance since they are a gauge invariant way of ``measuring'' the field
strength operator in either theory.

In the following we will partially identify a small set of non-chiral
operators.
Some of the identifications are possible due to degeneracies in the conformal
representation of these operators, allowing the calculation of their
conformal dimension (the identification is up to a certain well
defined ambiguity).

The identification is the simplest when assuming a canonical K\" ahler
potential, but is valid for an arbitrary one. In the general case
one sees that the exchanged operators correspond to geometrical
objects on the two corresponding K\" ahler manifolds.

In section 2 we discuss some gauge fixing conventions, most of which
are already in the literature, and some new ways of handling them
concisely.
Section 3 discusses the actual
correspondence. Section 4 contains a discussion of the correspondence
for SQCD with the addition of a matter field in the adjoint
representation.
Section 5 discusses the generalization to an arbitrary
K\"ahler potential. We conclude with a brief observation regarding the
gauge field strength in the theories.

\newsec{Notation and Equations of Motion}

We do not really have control over the non-chiral objects in a
strongly coupled supersymmetric field theory. But, following examples
from two
dimensional field theories, there are still interesting non-chiral objects
one can study.

The main example in two dimensional field theory is the correspondence
between
Landau-Ginzburg theories and minimal models, which was suggested in
\multic\ for non Supersymmetric theories (and thus has nothing to do with
chirality). We will therefore be
interested in essentially formal strings of the
allowed symbols at hand, i.e. fields and covariant derivatives,
modulo some relations. The relations are kinematical, such as the
constraints in the curvature tensor, and dynamical, such as the
equations of motion. Admittedly, this approach is not rigorously
justified and in fact can be shown to be wrong in some cases (for
a detailed analysis of an example see \nks)
but in this case it yields correct results as one can see from
assessing
the quantum corrections to the formula, at least
when the theory is taken to weak coupling. One can also compare it with
results from the representation theory of the superconformal algebra.

It is convenient to work in a covariant formulation \cov\covb\ in which gauge
transformations are taken to be maps from superspace into unitary
operators rather then maps of the form $e^{i\Lambda}$, where $\Lambda$
is a Lie algebra valued chiral superfield, which is the more commonly
encountered gauge.

More precisely, gauge transformations are given by elements in the
``group'', i.e. $e^{iK^iT_i}$ where $K$ are real superfields (and so the
lowest component of this transformation is a usual gauge
transformation). One also introduces a superspace gauge connection, which
is restricted by the equations
\eqn\curvconst{F_{\alpha\beta}=F_{\dot\alpha\dot\beta}=F_{\alpha\dot\beta}=0}
 where $F$ is the
superspace curvature form associated with the connection (these
equations are made more explicit below).
 One can pick the solution to
the constraints on the curvature form to be
$\sdd{\alpha}=e^{-\bar\Omega}D_{\dot\alpha}e^{\bar\Omega}$ and
$\sd{\alpha}=e^{-\Omega}D_{\alpha}e^{\Omega}$
 where $D$
is the familiar superspace covariant derivative and ${\bf D}$ is the gauged
covariant derivative (under an
infinitesimal gauge transformation the fields transform according to
$\delta\Omega\rightarrow
-iK+\bar\Lambda$, $\delta\bar\Omega\rightarrow -iK+\Lambda$). To pass
to the usual formulation one can
now choose a gauge in which $\Omega=-\bar\Omega$, and the usual gauge
field is $e^V=e^\Omega e^{-\bar\Omega}$.

Chiral fields in this formulation are defined as $\sdd{\alpha}\phi=0$
and are related to the usual definition by multiplying by appropriate
powers of $e^{\bar\Omega}$, and similarly antichiral fields will be
related to the usual definition by $e^\Omega$. From a more practical
point of view the formulas we will write below in this gauge will be
related to formulas in the usual gauge essentially by inserting $e^V$
whenever necessary, for
example $Q^\dagger e^V Q$ in the usual notation will now be just
$Q^\dagger Q$.

The constraints on the classical chain of symbols are kinematical and
dynamical. The kinematical constraints \curvconst\ on the curvature tensor are
$$\{\sd{\alpha},\sd{\beta}\}=\{\sdd{\alpha},\sdd{\beta}\}=0,\ \
{\bf D}_\mu\propto\{\sd{\alpha},\sdd{\beta}\}.$$
We will take the third one into account just by always writing ${\bf
D}_\mu$ in terms of $\sd{\alpha},\sdd{\alpha}$.

The dynamical constraints are the equations of motion. These are the
F-term equations
\eqn\eqmf{D^2Q=-4{\partial W^*\over\partial Q^*},\ {\bar D}^2Q^*=
-4{\partial W\over\partial Q}}
and the D-term equations, which for an $SU(N_c)$ theory with $Q$
($\widetilde Q$) fields in the
(anti) fundamental rep. are
\eqn\eqmd{{1\over g^2}D^{\alpha}W_{\alpha}=Q_iQ^{\dag}_i-{\widetilde
Q}_{\tilde i}^*{\widetilde Q}^T_{\tilde i}-{I_{N_c\times N_c}\over
N_c}
(Q_i^{\dagger} Q_i-{\widetilde
 Q}^T_{\tilde i}{\widetilde Q}^*_{\tilde i})}
rewriting the field strength superfield $W_{\alpha}$ in terms of the
curvature form \cov\wssbg, we obtain, in our notation, the relation
$${1\over g^2}\epsilon^{\beta\gamma}\epsilon^{\dot\delta\dot\beta}
(\sd{\gamma}\sd{\beta}\sdd{\beta}\sdd{\delta}-\sdd{\beta}\sdd{\delta}
\sd{\gamma}\sd{\beta}+2\sd{\gamma}\sdd{\beta}\sd{\beta}\sdd{\delta}
-2\sdd{\delta}\sd{\beta}\sdd{\beta}\sd{\gamma})=$$
\eqn\eqmc{=Q_iQ^{\dag}_i-{\widetilde Q}_{\tilde i}^*{\widetilde
Q}^T_{\tilde i}-{I_{N_c\times N_c}\over
N_c}(Q_i^{\dag} Q_i-{\widetilde Q}^T_{\tilde i}{\widetilde
Q}^*_{\tilde i})}

If one has a general K\" ahler form then the equations are modified and
are:

\noindent F-term equation:\eqn\eqmff{D^2 {\partial K\over\partial
Q^*}=-4
{\partial
W^*\over \partial Q^*}} and its conjugate.

D-term equation: we have to replace the above Yang-Mills current by an
expression that takes into account the K\"ahler form. The general
expression for a current, labeled by $a$, that corresponds to a
symmetry generated by a vector
field $v^{(a)}+v^{(a)*}$ on the K\" ahler manifold, is just
$v^{(a)}K$. Thus in our case ${1\over
g^2}D^{\alpha}W_{\alpha}^a=v^{(a)}K$, $a$ here is an index in the
adjoint of the gauge group. Both the $F$ and $D$ term
equations are invariant under K\"ahler transformations by a gauge
invariant holomorphic function.

A useful property of a symmetry current is \eqn\eqcc{{\bar D}^2(v^aK)={\bar
D}^2v^{a,i}{\partial K\over \partial x^i}=v^{a,i}{\bar D}^2{\partial
K\over\partial x^i}=-4v^{a,i}{\partial W\over\partial x^i}=0.} Note
that
if we assume that the global symmetry
currents are primary,
i.e. ${\bar S}_{\dot\alpha} (v^{(a)}K)=0$, then we consistently get from
the superconformal algebra that they have the correct dimension 2.

The operators that will be identified below live in degenerate
representations of the superconformal algebra. Following
\sucnfa\sucnfb, the representations of the N=1 superconformal algebra
are labeled by $(j,\bar j, D, R)$ where $j$($\bar j$) is the left
handed (right handed) $SU(2)$ representation, $D$ is the conformal
dimension (normalized to 1 on free scalar fields) and $R$ is the $R$
charge (normalized to 1 on gauginos). The representations fall into
three categories. Two of these categories are the chiral and anti-chiral
representations, which are the representations in which (in the
spinless case) the chiral (and anti-chiral) ring lives. We will be
interested here in the third type of representation. This type
satisfies $$D\ge\vert {3\over 2}R-j+\bar j\vert+j+\bar j+2$$ and it
becomes free when $j\bar j\not=0, {3\over 2}R=j-{\bar j}$ and $D=j+{\bar
j}+2$.

The current noted above is in this type of a representation with
$j={\bar j}=0$ and the representation is degenerate, satisfying $D=2$.

The other operators we will discuss below transform as $({1\over2},0)$
under the Lorentz group, their $(0,0)$ descendant is null and they satisfy
${3\over2}R<{1\over2}$. Thus their dimension satisfies $D=-{3\over
2}R+3$. If we now multiply such an operator by an anti-chiral operator,
then by standard arguments \sem\vafaa\ there are no short distance
singularities and result is again a degenerate
$({1\over2},0)$ operator, with the same kind of degeneracy, satisfying
${3\over2}R<{1\over2}$. Thus these operators form a module of
the anti-chiral ring (there is similarly a module of the chiral ring).

\newsec{SQCD and its dual}

Our main purpose is to identify some non-chiral operators between the
electric and the magnetic descriptions of SQCD.

For completeness, we will write down the electric and magnetic
theories. The electric theory contains fundamentals and
anti-fundamentals of $SU(n_c)$ with global charges
$$\vbox{\settabs 5 \columns
\+    &$SU(n_F)$ &$SU(n_{\bar F})$ &${U(1)}_B$ &${U(1)}_R$\cr
\+$Q$ &$n_F$     &$1$              &$1$        &$1-{n_c\over n_f}$\cr
\+$\widetilde Q$ &$1$  &$n_{\bar F}$     &$-1$       &$1-{n_c\over n_f}$\cr}$$
and no superpotential.

The magnetic theory contains fundamentals ($q$) and anti-fundamentals
(${\tilde q}$) of a
gauge group $SU(n_F-n_c)$ and an uncharged meson $M$, with global charges
$$\vbox{\settabs 5 \columns
\+    &$SU(n_F)$&$SU(n_{\bar F})$ &${U(1)}_B$ &${U(1)}_R$\cr
\+$q$ &${\bar n}_F$     &$1$       &$n_c\over n_F-n_c$ &$n_c\over n_f$\cr
\+$\widetilde q$&$1$&${\bar n}_{\bar F}$&$n_c\over n_F-n_c$ &$n_c\over n_f$\cr
\+$M$ &$n_F$     &$n_{\bar F}$     &$1$        &$2-{2n_c\over n_f}$\cr}$$
The superpotential in the magnetic theory is $cq{\bar q} M$ and $M$ is
identified with the electric composite $Q\bar Q$.

In this section we will take the K\" ahler potential in both theories to
be of the canonical form.

\medskip
The most immediate identifications are those of the global symmetry
currents. Generally:
\eqn\symmif{v^{(a)}_eK_e\leftrightarrow v^{(a)}_mK_m} for any symmetry with a
vector field generator $v^{(a)}$ on the K\" ahler manifold. The argument
for
these identifications is that the global symmetry current is a unique
object and it does not matter which variables we use to describe it.

In our case:
$$\vbox{\settabs 3\columns
\+$Current$ &$Electric$ &$Magnetic$\cr
\+$SU(n_F)$ &$P_T(Q^{iT}Q^*_j)$
&$P_T(-q_j^Tq^{i*}+M^{i\tilde k}M^*_{j\tilde k})$ \cr
\+$SU(n_{\tilde F})$ &$P_T({\widetilde Q}^{{\tilde i}T}{\widetilde
Q}^*_
{\tilde j})$ &$P_T(-{\tilde q}_{\tilde j}^T{\tilde q}^{{\tilde i}*}+M^{k\tilde
i}M^*_{k\tilde j})$ \cr
\+${U(1)}_B$ &$Q^{iT}Q^*_i-{\widetilde Q}^{{\tilde i}T}{\widetilde Q}_
{\tilde i}$ &${n_c\over n_f-n_c}(q^T_iq^{i*}-{\tilde q}^T_{\tilde i}
{\tilde q}^{{\tilde i}*})$ \cr }$$
where $P_T$ is a projection on traceless matrices. We will denote these
currents by $T,\ {\widetilde T}$ and $T_B$.

The two other identifications we would like to show are:
$$Electric \hskip 2in Magnetic \hskip 1.5in $$
\eqn\ida{{\widetilde Q}^T_{\tilde
k}\sd{\alpha}Q_i-{(\sd{\alpha}{\widetilde Q}_{\tilde k})}^TQ_i\ \
\leftrightarrow\ \ \epsilon^{\dot\alpha\dot\beta} (\bqd{k}
\sdd{\alpha}\sd{\alpha} \sdd{\dot\beta}\mqs{l})+products+descendant}
\eqn\idb{\epsilon^{\dot\alpha\dot\beta}{\widetilde Q}^\dagger_{\tilde k}
\sdd{\alpha}\sd{\alpha}\sdd{\beta} Q^*_i+({\widetilde Q}
\leftrightarrow Q)\ \ \leftrightarrow \ \ {\tilde q}^T_{\tilde k}
\sd{\alpha}q_l-{(\sd{\alpha}{\tilde q}_{\tilde
k})}^Tq_l}

These four operators are in a $({1\over 2},0)$ representation of the
Lorentz group. They have the property that, using the $D$ or the $F$
equations of motion, their $(0,0)$ descendant is either a product of
simpler operators or is null, and thus can be identified in the dual.
We see that
duality interchanges factorization via the $F$ equation of motion and
via the
$D$ equation of motion. Of course, the identification is up to
operators that have null $(0,0)$ descendant, a point to which we will
return later.

To show the equivalence, one proceeds as follows. Since one of the
theories in question is strongly coupled, we can not really do any
equivalent computation in both theories. What we will do, in the
spirit explained
above, is a classical calculation and then try to check whether it is valid
quantum mechanically.

We begin with \ida. The exact correspondence is:
\eqn\idafl{O^e={\widetilde Q}^T_{\tilde k}\sd{\alpha}Q_l-{(\sd{\alpha}{\tilde
Q}_{\tilde k})}^TQ_l}
maps to \eqn\idafll{O^m=\epsilon^{\dot\alpha\dot\beta} (\bqd{k}
\sdd{\alpha}\sd{\alpha} \sdd{\beta}\mqs{l}+{(\sdd{\beta}\sd{\alpha}
\sdd{\alpha}{\bqs{k}})}^T\mqs{l})+$$ $$+{1\over2}(
(\bqd{k}\sdd{\alpha}^2\mqs{l})-{(\sdd{\alpha}^2\bqs{k})}^T\mqs{l}))-$$
$$-{g^2\over 2c}(\sd{\alpha}(M_{s\tilde k}T^s_l)-2M_{s\tilde
k}(\sd{\alpha}T^s_l)$$ $$+{g^2\over 2c}(\sd{\alpha}({\widetilde T}^{\tilde
l}_{\tilde k}M_{l\tilde l})-2(\sd{\alpha}{\widetilde T}^{\tilde l}_{\tilde
k})M_{l\tilde l})+$$ $$+{g^2\over c}{n_F-n_c\over n_c}({1\over
n_f}-{1\over n_c})(\sd{\alpha}(M_{l\tilde k}T_B)-2M_{l\tilde
k}\sd{\alpha}T_B)+$$ $$+{g^2\over 2c}((\sd\alpha M^T)M^*M^T-M^TM^*
\sd{\alpha} M^T)}
And a direct classical calculation yields that the
$(0,0)$ descendant of both sides vanishes (and that all global
quantum numbers match).

By the general properties of the
representations of the superconformal algebra we know that the
conformal dimension of these operators is ${3\over2}+{3N_c\over
N_f}$. There are two ways
in which we can go to weak coupling. We can take $N_f\rightarrow 3N_c$
and approach weak electric coupling \bnks, in which case $D\rightarrow
{5\over2}$ which is the correct classical result for the electric
operator, or we can take $N_f\rightarrow {3\over2}N_c$
and approach weak magnetic coupling, in which case $D\rightarrow
{7\over2}$, which is the correct classical result for the magnetic
operator.

The fact that in the weak coupling limit the quantum dimension
converges to the classical dimension for this operator lends further
credence to the assumptions going into calculating its quantum
dimension, namely that it is primary and has a null descendant, and
that one should look for it in the dual theory.

We can obviously always add
to either sides descendants, or ``products'' of operators, with
appropriate quantum
numbers, that satisfy $D^{\alpha}O_{\alpha}=0$. We would like, however,
to claim that there is no arbitrariness in the part that contains the
covariant derivatives.

More precisely, we would like to show that there is no other operator
of the form, say, ${\widetilde Q}^T_{\tilde k}\sd{}\ldots \sd{} Q_l$ that has
the same
quantum numbers as ${\widetilde Q}^T_{\tilde k} \sd{\alpha} Q_l$ and whose
(0,0) $D_{\gamma}$ descendant is null. The reason is the following. If
the descendant is null and we know the R-charge then we
can calculate the conformal dimension of the operator to be
${3\over2}R+3$.
If we now take
the limit $N_c,N_f\rightarrow\infty$ at a fixed ratio ${N_f\over
N_c}\rightarrow 3-\epsilon$ then we are in a
weakly coupled theory and we can approximate the dimension of the operator
by the
classical dimension. In
particular each $D$ contributes ${1\over2}$ to the dimension. We thus
have the sum and difference (from the R charge) of the number of $D$'s
and $\bar D$'s and can easily see that
${\widetilde Q}DQ$ is the only allowed operator with such a degeneracy
for these quantum numbers.

Similarly, one is unable to construct any other operator of the form
${\tilde q}^\dagger D\ldots D q+descendants+products$ with the
required properties besides ${\tilde q}^\dagger {\bar D}
D_{\alpha}\bar D q$. The reasoning is similar.

Thus only these combinations of covariant derivatives have a chance of
living in a degenerate rep. of the Superconformal
algebra with $({1\over2},0)$ primary field, and indeed we have shown
that they can be extended to a degenerate
operator.

We cannot show that our manipulations hold Quantum Mechanically.
However, the picture presented thus far is the simplest, if not the
only, one that might work and that is self consistent.

To show \idb, we would like to calculate the $(0,0)$ descendants of both sides.
In the magnetic theory:
\eqn\idam{{-1\over 4c}\epsilon^{\rho\alpha} D_{\rho}O^m_\alpha={\tilde
q}^T_{\tilde k}{\tilde q}^{{\tilde i}*}M_{l\tilde i}^*-M_{{\tilde
s}k}^*q^{s\dagger} q_l=$$ $$=T^s_l M^*_{s\tilde k}-{\widetilde T}^{\tilde
i}_{\tilde k}M^*_{l\tilde i}-{n_f-n_c\over n_fn_c}T_B M_{l\tilde k}}
and in the electric theory:
\eqn\idae{\epsilon^{\rho\alpha} O^e_\alpha \propto
({\widetilde Q}^\dagger_{\tilde k} {\widetilde Q}_{\tilde
i})({\widetilde Q}_{\tilde i}^\dagger Q^*_{\tilde l})-({\widetilde
Q}^\dagger_k Q^*_j)(Q_j^T Q^*_l) -{1\over n_c}M_{l\tilde k}T_B=$$
$$= -(T^s_l) M^*_{s\tilde k}+{\widetilde T}^{\tilde
i}_{\tilde k}M^*_{l\tilde i}+({1\over n_c}-{1\over n_f})T_B M_{l\tilde k}}

Again, these are in agreement.

\newsec{Another Example}

The discussion above can be repeated in other models. An interesting
model has $SU(N_c)$ gauge group with $N_f$ quarks ($Q$) and anti-quarks
($\widetilde Q$) and a
matter field ($X$) in the adjoint representation of the gauge
group \kts\nks\ (and
some extensions of it in\otr). We will not repeat here the details of
the model.

In this model, if the electric superpotential is $Tr X^{k+1}$, then
one can define the degenerate electric operators ${\widetilde
Q}X^{*t}\sd{\alpha}Q-({\widetilde Q}\leftrightarrow Q)$ for $t<k$.
These are matched with a linear
combination of $\{{\tilde q}^*Y^{*s}\sdd{\ }\sd{\alpha}\sdd{\
}Y^{*k-1+t-s}q^*,s=1\ldots k-1+t\}$ (plus some descendants and products).

It is interesting to note the following relationship between the
degenerate operators that we have discussed, and the structure of the
chiral ring. Suppose we have an electric operator of the form
$C_1\sd{\alpha}C_2$ where
$C_1$ and $C_2$ are chiral operators and its magnetic counterpart is
$\epsilon^{{\dot\alpha}{\dot\beta}}{\hat C}_1^*\sdd{\alpha}\sd{\alpha}
\sdd{\beta}{\hat
C}_2^*$. It is then natural to have a term $M{\hat
C}_1{\hat C}_2$ in the magnetic superpotential, where $M$ is a
fundamental field in the magnetic theory that is identified with
$C_1C_2$ of the electric theory. Thus the product ${\hat C}_1{\hat
C}_2$
is zero
in the chiral ring, while ${\hat C}_1$ and ${\hat C}_2$ are not zero
in the chiral ring in the sense that there are non trivial chiral
primaries that contain them.

This is the case for the operator ${\widetilde
Q}\sd{\alpha}Q-(\sd{\alpha}{\widetilde Q})Q$. But we can also analyze
the operator ${\widetilde Q}X^*Q$ , in say the k=2 model. The
corresponding anti-chiral operator, after dropping the covariant
derivatives, is
${\tilde q}^*Y^{*2}q^*$. This operator can be simplified by the equation
of motion coming from $W_{mag}=TrY^3$, and is thus a sum of products
of simpler chiral primaries. Note that this operator could not have been
eliminated from the chiral ring by coupling it, in the superpotential,

to a magnetic gauge
invariant field. The operator it can couple to from the electric
theory is ${\widetilde Q}X^*Q$, but this
coupling not permitted in the superpotential since ${\widetilde
Q}X^*Q$ is not a chiral operator and therefore its magnetic
counterpart
cannot appear in the
superpotential at all.

Thus we see that in the examples above, the chiral operators obtained
in this way from the non-chiral module are products of non-trivial
operators such that the product factorizes in the chiral ring. Whether

this is a general feature or not is unclear,
nor is the part it may play in the larger picture.

\newsec{Correspondence with an Arbitrary K\"ahler Potential}

Since we do not know the effective K\"ahler potential that should be
used in these formulas, we would like to find
the
analogue of these correspondences for an arbitrary K\"ahler
potential,
again taking the classical objects of symbols modulo relations.

The first thing to note is that there is a preferred K\"ahler potential
in this problem, i.e. one can overcome the usually inherent ambiguity
associated with K\"ahler transformations. Given an arbitrary K\"ahler
potential it is natural to
define the invariant K\"ahler potential (obtained by averaging over
orbits of all the symmetries in the problem, and particularly the
R-symmetry). Once done, then the inherent
arbitrariness
in K\"ahler transformations $K\rightarrow K +F(X)+F^*(X^*)$ is gone
since, by unitarity constraints,  there
is no gauge invariant holomorphic regular function with R charge 0.

The generalization of ${\tilde q}Dq-(D{\tilde q})q$ is $O_{\dot\alpha}
=\kqbs{m}{i}{k}
({\sd{\alpha}}\kqs{m}{i}{l})-(\sd{\alpha}\kqbs{m}{i}{k})\kqs{m}{i}{l}$.
Note that the matching of quantum numbers requires that $K_m$ be the
invariant K\"ahler potential above. Then
$\epsilon^{\dot\beta\dot\alpha}\sdd{\beta}O_{\dot\alpha}=
-4(\kqbs{m}{i}{k}\wqs{m}{i}{l}-\wqs{m}{i}{l}\kqbs{m}{i}{k})$, which
can be written in terms of the global symmetry currents. Similarly in
the electric theory the (0,0) descendant is still
null because $W=0$.

There are two points to be made about this operator. First is that it
can be generalized in a trivial way
$O_{\dot\alpha}^{(P)}=\kqbs{m}{i}{k}P(X)({\sdd{\alpha}}\kqs{m}{i}{l}-
(\sdd{\alpha}\kqbs{m}{i}{k})P(X)\kqs{m}{i}{l}$
where $P$ is any holomorphic polynomial in whatever fields we have in
the theory.

The second comment regards the geometrical meaning of
$O_{\dot\alpha,{\bar k}l}$. Let us discuss for a minute a similar
object in a WZ model, i.e., the object $({\partial K\over \partial
X^{i*}}(D_{\alpha}{\partial K\over\partial
X^{j*}})-(D_{\alpha}{\partial K\over\partial X^{i*}}){\partial
K\over\partial X^{j*}}){dx}^{i*}\wedge{dx}^{j*}$. This is clearly a closed
two form. The transformation law of
our object $O$ above is similar but it is not quite a 2 form. We can however
write it as the components of the two form
$(\kqbs{m}{i}{k}(\sdd{\alpha}\kqs{m}{j}{l})-(\sdd{\alpha}\kqbs{m}{i}{k})
\kqs{m}{j}{l})dq^{jl*}d{\tilde
q}^{*i\tilde k}$ evaluated on an appropriate set of gauge invariant
bi-vectors.

The second object, of the type ${\tilde Q}^*\sdd{\ }\sd{\ }\sdd{\ }Q^*$,
remains the same in the theory with an arbitrary K\"ahler potential.
The only change is that one has to insert the gauge
current appropriate to K. Expression 3.5 remains essentially the
same except that every $M$ appearing should be replaced by ${\partial
K\over \partial M^*}$.

\newsec{Final Comment}

We would make one final observation. Calculating the (1,0) descendant
of both sides of (3.3), one obtains
\eqn\veryimp{{\widetilde Q}^*F_{\alpha\beta}Q^*\leftrightarrow
\psi^m_{\{\alpha}{\tilde\psi}^m_{\beta\}}}
where $\psi^m({\tilde\psi}^m)$ are the fermion partners of $q$ and
${\tilde q}$.

The physics behind this expression is unclear. It is intriguing,
however, that at the end of the day the field strength (as measured in
the LHS) is a fermion matter bi-linear in the dual theory.

\centerline{{\bf Acknowledgments}}

It is a pleasure to thank P. Argyres, T. Banks, M. Peskin and
N. Seiberg for useful discussions and comments, and to thank SLAC for
its hospitality at the early stages of this work.

\listrefs
\end